# UV Irradiation Facility for Solar Effects Simulations

A.B. Alpat[1], G. Bartolini, S. Bollanti, P. Di Lazzaro, D. Murra, T. Wusimanjiang

*Abstract*— We describe an experimental setup developed aiming to irradiate samples under UV radiation (200 nm-400 nm) for accelerated test for solar effects according the relevant ECSS/ESA standards. This facility has been already used for projects belonging to large space programs (Cosmic Vision, Artes) for simulations up to 3,500 e.s.h. (equivalent sun hours). In particular, we detail the calculation of the UV dose delivered by Sun, the calibration of the detectors, the spatial distribution of the UV radiation on samples, the remote control 24/24-7/7 of both samples' temperature and lamp radiation, the sample's heat dissipation and operation in a helium atmosphere.

## I. INTRODUCTION

The Sun irradiates charged particles (protons, electrons, heavy nuclei) as well as high-energy photons at near and vacuum ultraviolet (NUV and VUV) radiation, which can damage space devices orbiting out the atmosphere. These particles and UV radiation can cause both hardening and weakening of polymers that compromise elements of the vehicle itself or its payloads, degrade optics and windows, destroy electronics, and pose serious threat to astronauts [1]. The impacts of the solar environment vary greatly, depending on distance from the Sun. Spacecraft scattered across the solar system, whether in Earth orbit or journeying to outer planets or approaching the Sun itself, experience dramatically varying conditions. Whatever the destination, spacecraft and crews must have reliable protection from the threats the Sun poses, with the best materials available. The principles of the particle and electromagnetic radiation test of space materials is to evaluate their physical properties changes under specific laboratory simulations that imply a suitable simplification of space degrading factors, and, generally, a reduction of the irradiation time. Such approach is performed following the major space agency standards, in our case, the ECSS standard given in [2]. Examples of the main properties used in space technology are transmission/absorption of windows, thermo-optical properties (spectral and solar absorptance, infrared emissivity) and electrical properties.

The Sun emits radiation from X-rays to radio waves, but the irradiance of solar radiation peaks in the visible wavelengths (see Fig. 1). Common units of irradiance are Joules per second per $m^2$ of illuminated surface per nm of wavelength (e.g., between 200 nm and 400 nm), i.e. W $m^{-2}$ $nm^{-1}$ as in the plot in Fig.1. To get the total irradiance in units of W $m^{-2}$, the spectral irradiance should be integrated over the wavelength interval of interest.

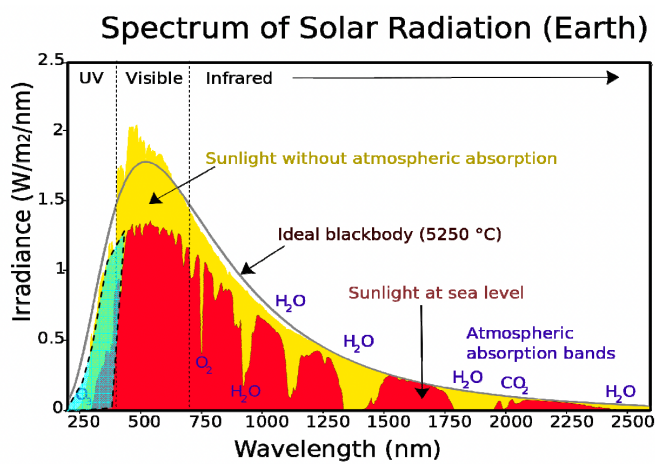

Fig. 1. Solar irradiation spectrum. The green shaded area is that of 200nm-400nm (Credit: CC BY-SA 3.0 via Wikimedia Commons)

In the range between 200 and 400 nm (see [3] section 4.2.6 therein), the solar power density is about 118 W/$m^2$. This is the UV irradiance out of the atmosphere (in low Earth orbits).

## II. IRRADIATION FACILITY AND EXPERIMENTAL SETUP

The test facility is located at ENEA Research Center, in Frascati. The experimental setup has been developed in cooperation between BEAMIDE srl [4], INFN Sezione di Perugia [5] and ENEA-Frascati [6]. The samples are placed on the bottom of a cylindrical vacuum chamber equipped with a water-cooling system running between double-sleeve inox walls and closed on the top by a quartz window. A Helios Quartz mercury lamp (Polymer model, see www.heliosquartz.com) was placed above the chamber as UV source. The chamber, after a short degassing period (vacuum at $10^{-4}$ mbar), was filled with 1.1 bar of He in closed recirculation

---

[1] A.B. Alpat is with Istituto Nazionale di Fisica Nucleare, Sezione di Perugia, Via A.Pascoli snc, 06123, Perugia, Italy (e-mail: behcet.alpat@pg.infn.it). G. Bartolini and T. Wusimanjiang are with BEAMIDE s.r.l., Via Campo di Marte 4/o, 06124, Perugia, Italy (e-mails: giovanni.bartolini@beamide.com and talipjan.osman@beamide.com),

S.Bollanti, P. Di Lazzaro and D. Murra are from ENEA Frascati Research Center, via Enrico Fermi 45, 00044, Frascati (RM), Italy, (emails: sarah.bollanti@enea.it, daniele.murra@enea.it, paolo.dilazzaro@enea.it).

with an oxygen sensor to monitor any possible contamination or contact of samples with air. The overall setup includes water and He flow cooling systems, a photodiode to monitor ultraviolet (UV) light, an analyzer of the percentage of oxygen present in the chamber, a data acquisition system and a webcam with remote control of the main experimental parameters (Fig. 2).

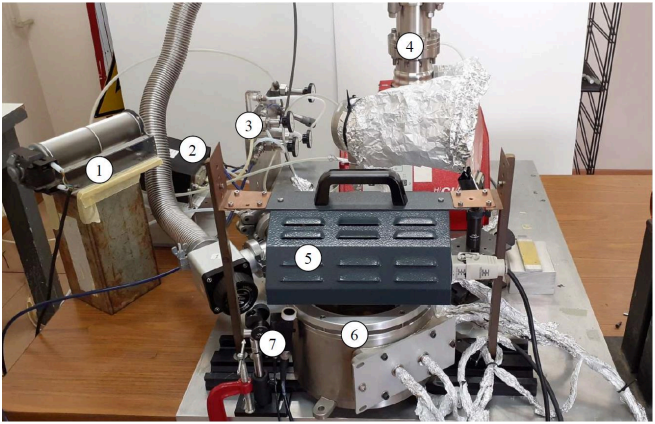

Fig. 2. Irradiation setup at the ENEA Research Center in Frascati. 1) tangential fan; 2) oxygen analyzer; 3) panel with taps and flow meter for controlling the gas circuit; 4) molecular turbo pump; 5) Helios Quartz lamp; 6) irradiation chamber with inlet / outlet for connections of the temperature sensor and cooling water; 7) Thorlabs photodiode for the monitoring of UV light.

In Fig. 2, the He gas recirculation system, the webcam and the data logger for the data acquisition are not visible. The hydraulic circuit of the helium gas recirculation, used for heat dissipation and for the control of residual oxygen, is shown in Fig. 3.

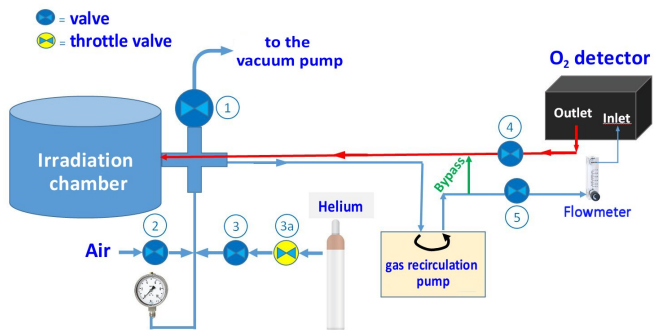

Fig. 3. Schematic diagram of the hydraulic circuit for helium gas flow circulation, which is used to keep the samples from being exposed to air and to promote heat dissipation associated with long term irradiation.

In order to keep the temperature of the samples under irradiation at a low and constant value it is important to avoid secondary effects. The inert gas recirculation with controlled pressure in the chamber provides further heat dissipation in addition to water cooling of the walls and avoids samples to get in direct contact with air. In all tests, we recorded the irradiated samples temperature that never exceeded 30 °C. The oxygen pressure in the chamber was always lower than few ppm.

The Helios Quartz lamp, like all similar lamps, emits spatially inhomogeneous radiation. Furthermore, the internal walls of the vacuum chamber reflect the incident rays with an effectiveness that depends on the angle of incidence of the radiation. In practice, the side walls of the chamber act as a "waveguide" improving the symmetry of the spatial distribution of the UV irradiance on the bottom of the chamber.

To evaluate as accurately as possible the incident irradiance on the exposed samples, a measure of the irradiance spatial distribution at the bottom of the chamber is needed. First of all, we measured the absolute irradiance value in the center of the chamber and in two other lateral points using the Hamamatsu absolute power meter (model C9536/H9535, see Fig. 4).

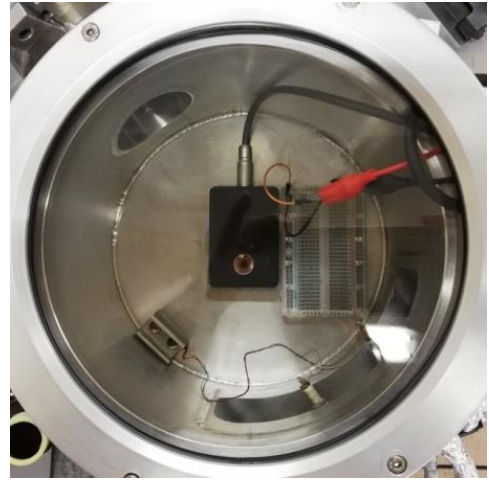

Fig. 4. A photo of the Hamamatsu absolute power-meter sensor at the bottom of the chamber for measuring lamp irradiance. Next to the sensor there is an experimental board with a phototransistor, for its calibration.

Then, since it was not possible to use the bulky power meter to have a map of the power density on the whole bottom of the chamber, we calibrated the response of a phototransistor (model PT334-6C of Everlight) to the light emitted by the lamp.

Finally, we realized a board with 17 phototransistors of the same type of the calibrated one, having previously checked that they had the same response characteristics. We placed the board on the bottom of the chamber, so as to know the irradiance distribution more accurately (see Fig. 5). This measurement is essential for attributing to each position assumed by the samples during the irradiation the correct irradiance and, consequently, the relationship between this irradiance and the solar one (which we will call acceleration factor FA). The acceleration factors provided in our system are in compliance with the requirements of [3] as reported in section 6.3.3.7 therein.

Once the measurement of the UV irradiance distribution on the bottom of the chamber is completed, the irradiation can start. This was divided in 4 phases to allow for a controlled exchange of the samples positions, and therefore to obtain a more homogeneous irradiation. Fig. 6 shows the placement of some polyamide-based metamaterial samples.

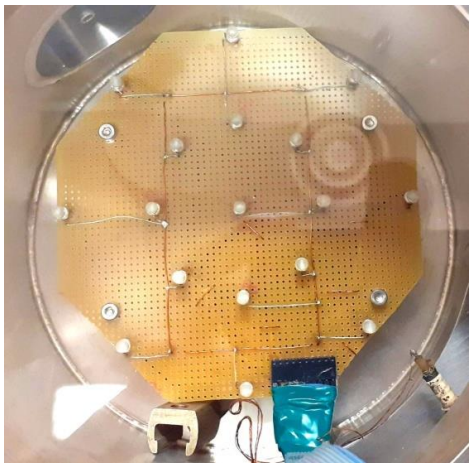

Fig. 5: The uniformity of the irradiation area was mapped at each phase of the irradiation session. The mapping was done through phototransistors placed over the irradiation area.

In order to calculate precisely the number of e.s.h. experienced by each sample, in addition to the relative irradiance between the various zones of the irradiation chamber, it is necessary to know the temporal evolution of the lamp power. This was possible thanks to the Thorlabs 201 photodiode, placed next to the chamber (see Fig. 2), equipped with a ZWB2 filter to select a portion of the spectrum centered in the ultraviolet.

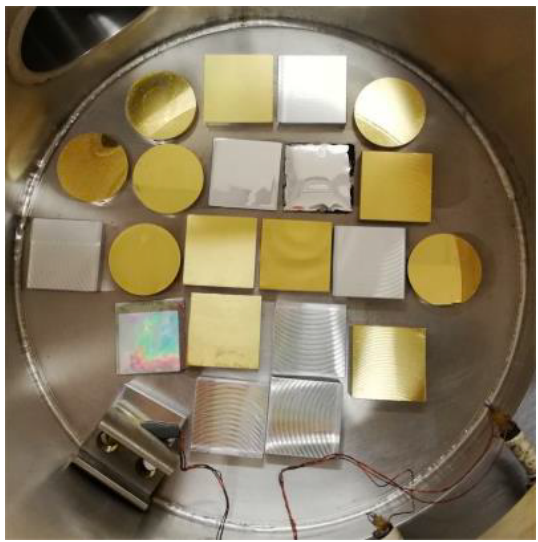

Fig. 6: Samples placed on the bottom of the 18-cm diameter irradiation chamber.

### III. Conclusion

We have developed a compact UV irradiation system at the ENEA Frascati Research Center Laboratories, where the described facility has been used for 200-400 nm UV testing of target materials for several ESA supported projects [6, 7]. The system is fully compliant with relevant ESA standard's requirements [2] and it is audited by ESA technical officers. The user can control the testing conditions (oxygen contamination, helium flow, ambient and sample temperature, elapsed time, total power delivered etc.) during any time of the irradiation phase.

### Acknowledgment

Authors would like to thank colleagues, S. Mengali e M. Simeoni from Consorzio CREO [8] for their valuable comments during setting up the experimental setup and for providing us the samples shown in Fig. 6.